\documentclass[11pt]{article}

\usepackage{amsmath}
\usepackage{jheppub}
\usepackage{amssymb}
\usepackage{graphicx}
\usepackage[T1]{fontenc}
\usepackage{slashed}
\usepackage[utf8]{inputenc}
\usepackage{xspace}
\usepackage{color}
\usepackage{url}
\usepackage{array}
\usepackage{ulem,fancyvrb,multirow}
\usepackage{xcolor}
\usepackage{mathtools}
\allowdisplaybreaks
\date{\today}

\begin{document}
%\linenumbers

\title{Constraints on real scalar inflation from preheating of LATTICEEASY}

\author[a,b,1]{Wei Cheng}
\author[a]{, Tong Qin}
\author[a]{, Jiujiang Jiang}
\author[a,b,1]{and Ruiyu Zhou}
\affiliation[a]{School of Science, Chongqing University of Posts and Telecommunications, Chongqing 400065, P. R. China}
\affiliation[b]{Department of Physics and Chongqing Key Laboratory for Strongly Coupled Physics, Chongqing University, Chongqing 401331, China}
\emailAdd{chengwei@cqupt.edu.cn}
\emailAdd{zhoury@cqupt.edu.cn}
\footnotetext[1]{Corresponding authors.}

\abstract{
In this paper, we undertake a detailed study of real scalar inflation by employing LATTICEEASY simulations to investigate preheating phenomena. Generally, the scalar inflation potential with non-minimal coupling can be approximated by a quartic potential. We observe that the evolutionary behavior of this potential remains unaffected by the coupling coefficient. Furthermore, the theoretical predictions for the scalar spectral index ($n_s$) and tensor-to-scalar power ratio (r) are also independent of this coefficient. Consequently, the coefficients of this model are not constrained by Planck observations.
Fortunately, the properties of preheating after inflation provide a viable approach to examine these coefficients. Through LATTICEEASY simulations, we trace the evolution of particle number density, scale factor, and energy density during the preheating process. Subsequently, we derive the parameters, such as the energy ratio ($\gamma$) and the e-folding number of preheating ($N_{pre}$), which facilitate further predictions of $n_s$ and $r$. We have successfully validated real scalar inflation model using preheating of LATTICEEASY simulations based on the analytical relationship between preheating and inflation models.}
\maketitle
\preprint{}

\section{Introduction}

The inflationary paradigm has emerged as an elegant solution to several challenges within the standard model of cosmology~\cite{Guth:1980zm,Linde:1981mu}, capturing the attention of numerous researchers since its proposal. Experimentally, precise measurements of the cosmic microwave background radiation temperature and anisotropic polarization provide strong support for the  slow-rolling inflation~\cite{Planck:2015bpv, Planck:2018jri}. Theoretical investigations have led to the proposal of numerous inflation models, with the single-field inflation model being the simplest and most widely studied. Examples of such models include polynomial potential models ~\cite{Martin:2013tda,Martin:2006rs,Martin:2010kz,Dai:2014jja}, the Starobinsky model \cite{Starobinsky:1980te,Jin:2020tmm,Mantziris:2022fuu}, Higgs Inflation \cite{Bezrukov:2007ep,Gundhi:2018wyz,Antoniadis:2021axu,Liu:2022myp,Barrie:2021mwi}, Natural Inflation~\cite{Freese:1990rb,Zhang:2018wbn,Odintsov:2019mlf,Cheng:2021qmc}, Hilltop Inflation ~\cite{Linde:1983psb,Bostan:2019wsd,Kallosh:2019jnl}, and others.

The testing of numerous inflation models has become a central challenge in the field of inflationary physics. Unfortunately, direct observations of inflation in the early universe are not possible, and researchers rely on indirect methods to test these models. One common approach is to utilize the scalar spectral index and tensor-to-scalar power ratio, which are predicted by Planck's observations of the cosmic microwave background (CMB). As the accuracy of Planck's observations has improved, it has led to the exclusion of many polynomial inflation models~\cite{Planck:2015bpv, Planck:2018jri}. However, a promising avenue to address this issue is to consider the non-minimal coupling between the scalar inflaton and the Ricci scalar. This framework offers a mitigation scheme that can potentially rescue a significant number of inflation models~\cite{Choubey:2017hsq,Lebedev:2021zdh,SBudhi:2019vln,Cheng:2022hcm}. By incorporating this non-minimal coupling, the problematic aspects of certain inflation models can be alleviated, providing a more favorable agreement with the observational data from Planck and preserving their viability.

Considerable efforts have been devoted to further constraining scalar inflation models by exploring the interconnectedness of various physical phenomena. After inflation ends, the universe enters a reheating stage~\cite{Albrecht:1982mp,Kofman:1994rk,Kofman:1997yn,Lozanov:2019jxc,Saha:2020bis}, during which the inflaton decays into standard model particles or even dark matter particles, and the universe will be heated. Thus the reheating may provide a platform to study the inflation.

The limitations of reheating in the context of the single-field scalar inflation model have been extensively studied in Ref.~\cite{Cook:2015vqa}. Reheating involves a preheating process during the early stages, as described in works such as~\cite{Kofman:1994rk,Kofman:1997yn,Greene:1997fu,Dolgov:1989us,Traschen:1990sw,Shtanov:1994ce}.
The preheating phase plays a crucial role as it enables efficient transfer of energy from the inflaton field to the coupled fields. This rapid, non-thermal process is vital for reheating the universe, establishing the initial conditions for the hot Big Bang phase. Without preheating, the transition from the inflationary phase to the radiation-dominated era would occur at a considerably slower pace and with lower efficiency, possibly resulting in discrepancies with observational cosmology. The presence of preheating can be discerned through its influence on the power spectrum of CMB fluctuations and the formation of large-scale structures. Specifically, models with preheating predict distinct non-Gaussianities in the CMB fluctuations, which can be probed by current and future observational missions.

A comprehensive analysis of the interplay between preheating and inflation is presented in Ref.~\cite{ElBourakadi:2021nyb}. Furthermore, the constraints imposed by preheating on $\alpha$-attractor inflation models are investigated by allowing for flexible parameter choices in the properties of the preheating stage, as discussed in Ref.~\cite{ElBourakadi:2022lqf}.
The complexity of the preheating process arises from several factors, including the non-perturbative generation of material particles and the rapid energy transfer. This complexity gives rise to highly nonlinear and uncertain physics, making the study of preheating challenging and necessitating careful consideration of its implications on inflationary models.

Many non-perturbative reheating models have been proposed to study the nature of preheating, which include the parametric resonance model~\cite{Kofman:1997yn,Cai:2021yvq}, the hyperluminal instability model~\cite{Felder:2000hj,Enqvist:2005qu,Dufaux:2008dn}, the instantaneous preheating model~\cite{Felder:1998vq,Panda:2009ji,Dimopoulos:2017tud} and the holographic preheating cosmological model~\cite{Cai:2016sdu,Cai:2016lqa}. In our study, we focus on investigating the inflation model by leveraging the properties of preheating. Specifically, we consider the minimal scalar inflation model, which is a simplified version derived from the Polynomial potentials model. To analyze the preheating stage, we employ the LATTICEEASY simulation framework~\cite{Felder:2000hq}, which enables us to replicate and study the evolution of preheating processes.

Specifically, we extensively investigated the evolution of particle number density, scale factor, and energy density during the preheating process using LATTICEEASY simulations. By analyzing the simulated data, we deduced important parameters such as the energy ratio $\gamma$ and the e-folding number of preheating $N_{pre}$. Additionally, we established an analytical relationship between the preheating process and the inflation model under study. By combining this analytical relation with the observed properties of preheating, we were able to derive constraints on the minimal scalar inflation model. These constraints provide valuable insights into the viability and parameter space of the inflation model, thereby enhancing our understanding of its dynamics and predictions.

The rest of this work is organized as follows.
In Sec.~\ref{sec:PCMod}, we briefly review the minimal scalar inflation model and preheating constraints, and study the S-field driven inflation in detail.
Then in Sec.~\ref{sec:LATTICE}, we make a simulation for the preheating by applying LATTICEEASY, and discuss the evolution of particle number density, scale factor, and energy density.
Numerical analysis and discussion are presented in Sec.~\ref{sec:Num}.
We make a summary in Sec.~\ref{sec:Sum}.

\section{Preheating constraints on real scalar inflation}\label{sec:PCMod}

\subsection{S-field driven inflation}\label{sec:Inf}

Considering that inflation is driven by the S-field, one can write the corresponding action as:
\begin{align}
S_J = \int d^4 x \sqrt{-g} \left[\frac{M_P^2}{2} R \left(1 + 2 \xi_S \frac{S^2}{M_P^2} \right) -\frac{1}{2} \left(\partial S \right)^2 - V(S) \right]\,,
\end{align}
Following the conformal transformation strategy~\cite{Kaiser:2010ps}, the action $S_E$ in the Einstein frame can be easily obtained as:
\begin{align}
S_E = \int d^4 x \sqrt{- \tilde{g}} \left[\frac{M_P^2}{2} \tilde{R} - \frac{1}{2 }  \left(\tilde{\partial} \bar{S} \right)^2 - V(\bar{S}) \right],
\end{align}
where the potential $V(\bar{S})$ is expressed as:
\begin{align}
V (\bar{S})= \frac{\lambda_S M_P^4}{16 \xi_S^2}  \left(1 - e^{- 2{\sqrt{\frac{2}{3}}} \frac{\bar{S}}{M_P}  } \right)^2,\label{eq:V}
\end{align}
and the relation between the refined $\bar{S}$-field and $S$-field is expressed as:
\begin{align}
\bar{S}= \sqrt{\frac{3}{8}} M_P \ln \left(1+ \frac{2\xi_S S^2}{M_P^2} \right).
\end{align}

\begin{figure}[!tp]
\begin{center}
\includegraphics[width=0.6\textwidth]{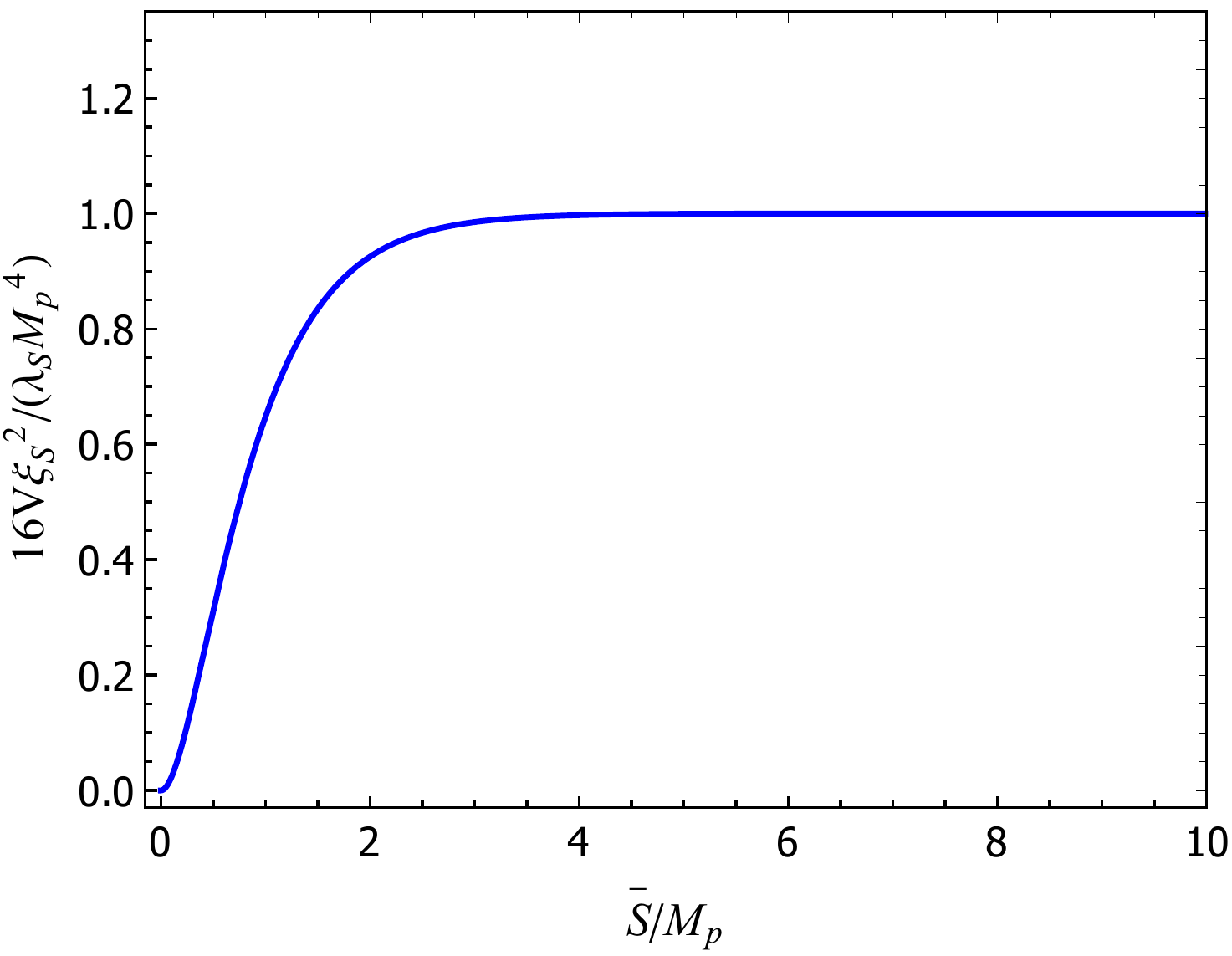}
\caption{The variation of slow-rolling inflationary potential with respect to inflaton, which calculated by using Eq.~(\ref{eq:V}). In the early stages of inflation, the scalar field rolls slowly in the direction that it fall. Then, when the potential energy is no longer dominant, inflation ends.}
\label{fig:V}
\end{center}
\end{figure}

The variation of potential $V$ with $\bar{S}$, shown in Fig.~\ref{fig:V}, indicates the trend of the potential is independent of the coefficient $\xi_S$ and $\lambda_S$. As $\bar{S}$ goes up, $V$ goes up and after some time, it reaches a plateau, which makes it possible to achieve slow-rolling inflation.

With potential at hand, one can study cosmological inflation in detail.
For the e-folding number between the horizon exit of the pivot scale and the end of inflation, it can be analytically calculated according to the following formula~\cite{Zhou:2022ovp}:
\begin{align}
N_k = \frac{1}{M_P^2} \int_{\bar{S}_{end}}^{\bar{S}_k} \frac{V}{V'}\, d \bar{S}=\frac{\sqrt{6}}{8 M_P} \left[\frac{\sqrt{6}M_P }{4}  e^{ 2{\sqrt{\frac{2}{3}}} \frac{\bar{S}}{M_P} } - \bar{S}  \right] \Big|^{\bar{S}_k}_{\bar{S}_{end}}.
\end{align}
Since $\bar{S}_k \gg \bar{S}_{end}$, and $\frac{\sqrt{6}M_P }{4} e^{\sqrt{\frac{2}{3}} \frac{\bar{S}_k}{M_P}} \gg \bar{S}_k$ , we have
\begin{align}
N_k = \frac{3}{16} e^{ 2{\sqrt{\frac{2}{3}}} \frac{{\bar{S}_k}}{M_P} }.
\end{align}
It means that
\begin{align}\label{eq:Sk}
\bar{S}_k = \frac{\sqrt{6}}{4} M_P \ln \left(\frac{16}{3} N_k \right)
\end{align}

According to the definition of slow-rolling parameters ($\epsilon = \frac{M_{\rm p}^2}{2} \left(\frac{dV/d\bar{S}}{V} \right)^2, \eta= M_{\rm p}^2 \left( \frac{d^2V/d\bar{S}^2}{V} \right)$), and combined with the Eq.~(\ref{eq:Sk}), the slow-rolling parameters can be obtained as follows:
\begin{align}\label{eq33}
\epsilon_k \simeq  \frac{3}{16N_k^2},\quad\quad\quad \eta_k = - \frac{1}{N_k}.
\end{align}

Further according to the relationship between the scalar spectral index $n_s = 1 - 6 \epsilon + 2 \eta$ (tensor-to-scalar power ratio $r=16 \epsilon$) and slow-rolling parameters, we can get:

\begin{align}
r &\simeq  \frac{3}{N_k^2},\\
n_s&\simeq 1- \frac{9}{8N_k^2}- \frac{2}{N_k}.\label{eq:ns}
\end{align}

Meanwhile, the amplitude of power spectrum of primordial power spectrum $A_s$ can be expressed as
\begin{align}
A_s=\frac{1}{24\pi^2M_p^4}\frac{V}{\epsilon}\Big|_{\bar{S}_{k}},
\label{eq:As}
\end{align}
The CMB observation indicates $A_s=2.2\times 10^{-9}$~\cite{Planck:2015sxf}.

%From the Fig.~\ref{Fig_ns_r.pdf}, one can find that these predictions fit very well with the latest PLANCK data for $N_k=50$ to $60$~\cite{Planck:2018jri}.

From Eq.~(\ref{eq:ns}) then one finds
\begin{align}\label{nk}
N_k = \frac{2}{(1 - n_s)}.
\end{align}

Finally, the $H_k$ and $V_{end}$ as a function of $n_s$ and $A_s$ can be derived by applying Eq.~(\ref{nk})~\cite{Cook:2015vqa}:
\begin{eqnarray}\label{n2}
& &H_k = \pi M_P  \sqrt{\frac{3}{2} A_s} (1- n_s),\\\label{n3}
& &V_{end} = \frac{9}{2} \pi^2 M_P^4 A_s (1-n_s)^2 \frac{ \left[\frac{2}{13} (4-\sqrt{3})\right]^2}{ \left[\frac{1}{64}(29+3n_s) \right]^2}
\end{eqnarray}
Eqs.~(\ref{nk}-\ref{n3}) are all the procedures to help derive the results for the preheating constraints on the minimal scalar inflation model.

\subsection{Preheating constraints}\label{sec:Inf}

When inflation ends, the universe becomes cold and empty, and the reheating process will heat the universe up to the temperatures required for Big Bang nucleosynthesis. Usually, at the beginning of reheating there is a period of explosive growth of particles called preheating~\cite{Lozanov:2019jxc}. Its duration is expressed as $N_{pre}$, which can be described analytically by~\cite{Koh:2018qcy, ElBourakadi:2022lqf}
\begin{equation}
N_{pre}=\left[ 61.6-\frac{1}{4}\ln \left( \frac{V_{end}}{\gamma H_{k}^{4}}%
\right) -N_k\right] -\frac{1-3\omega_{th} }{12(1+\omega_{th} )}\ln \left( \frac{3^{2}\cdot 5~V_{end}}{\gamma
\pi ^{2}\bar{g}_{\ast }T_{re}^{4}}\right),  \label{pre}
\end{equation}%
where $\gamma$ is the ratio of the energy density at the end of inflation to the preheating energy density, i.e, $\gamma=\rho_{pre}/\rho_{end}$~\cite{ElBourakadi:2021nyb}. Although the equation of states ($\omega_{th}$) here is for the reheating period, that of the preheating ($\omega_{pre}$) is numerically the same as $\omega_{th}$ in the reheating period, which can be obtained by deducing the relation between $\rho_{end}/\rho_{pre}$ ($\rho_{end}/\rho_{th}$) and $\omega$ respectively. Therefore, we will omit subscripts in subsequent discussions. The energy $\rho_{end}$ is related to the potentia at the end of inflation~\cite{Koh:2018qcy}:
\begin{equation}
\rho_{end}=\lambda_{end} V_{end},  \label{rend}
\end{equation}
where $\lambda_{end}=\frac{6}{6-2\epsilon}|_{S=S_{end}}=3/2$.

The Eq.~(\ref{pre}) entails a close connection between the preheating and the inflationary model, and the ambiguous preheating properties is a hindrance to the application of the inflationary model. By using LATTICEEASY to simulate the evolution of S- and h- fields in the preheating, we can infer the specific values of $N_{pre}$ and $\gamma$, which will be discussed in detail in the next section.

\section{Preheating in the LATTICEEASY}\label{sec:LATTICE}

As the physical properties of the preheating process are highly non-linear and complex, so the effective research method is usually related to lattice simulation. In this section, we will apply LATTICEEASY~\cite{Felder:2000hq} to reconstruct the physics of the preheating and test the scalar inflation model, which simulate the emergence of the Higgs field subsequent to the damping of the S-field through a set of coupled field motion equations.

Consider the following couplings as illustrative examples: $\lambda_S=10^{-13}$, $\lambda_{Sh}=2\times10^{-11}$ and $\lambda_{h}=8\times10^{-12}$. Firstly, the choice of $\lambda_{Sh}$ directly impacts the generation process of Higgs. When $\lambda_{Sh} \geq 10^{-11}$, the production of Higgs may occur through either the freeze-in or freeze-out mechanism, for which a more in-depth discussion can be found in Ref.~\cite{Lebedev:2021ixj}. Secondly, the parameters $\lambda_S$ and $\lambda_h$ represent the coupling strengths of the S field and the Higgs field, respectively, and they significantly impact the dynamics of the system. Ref.~\cite{Lebedev:2021zdh, Prokopec:1996rr} indicates that a strong self-interaction of the Higgs field can lead to a large effective mass term, thereby suppressing resonance phenomena. Therefore, the selection of $\lambda_S$ and $\lambda_h$ is crucial for the outcomes of our simulations. In our simulation strategy, we deliberately chose smaller values for $\lambda_S$ and $\lambda_h$. This choice results in a smaller effective mass term, which in turn enhances the resonance phenomena, thereby significantly enhancing the backreaction effect during the preheating process. This approach allows us to deeply explore the significance of backreaction effects during the preheating phase and their impact on particle dynamics under early universe conditions.
We will discuss in detail how to use preheating to constrain the inflation model~\cite{Lebedev:2019ton, Lebedev:2021zdh}.

\begin{figure}[!htp]
\begin{center}
\includegraphics[width=0.55\textwidth]{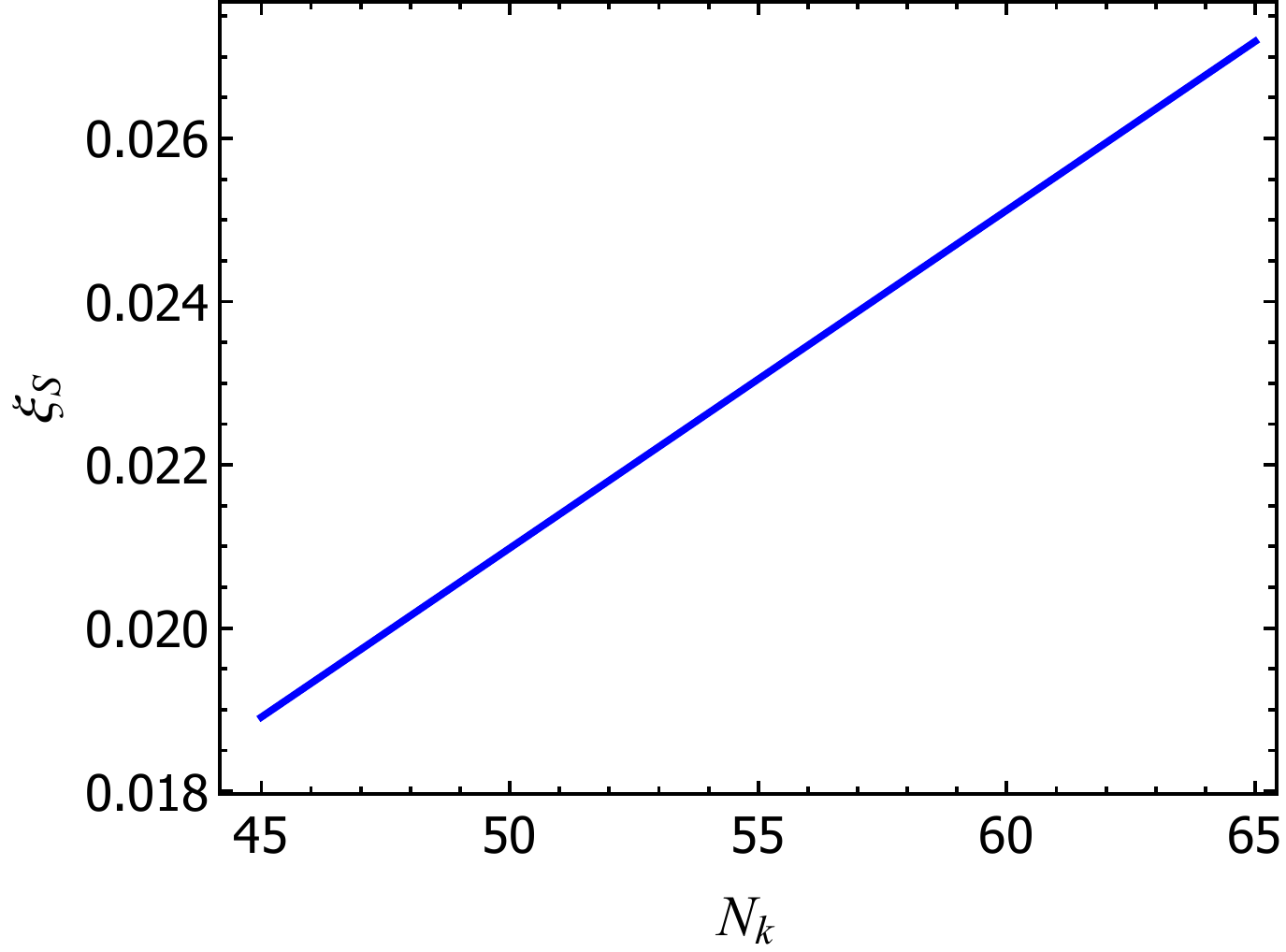}
\caption{The non-minimal coupling parameter $\xi_s$, as determined by Eq.~(\ref{eq:As}) with $\lambda_S=10^{-13}$, varies as a function of $N_k$. The parameter $\xi_s$ shows a rough proportionality to $N_k$, yet it experiences negligible variation despite changes in $N_k$.}
\label{fig:Nk_xis}
\end{center}
\end{figure}

Fig.~\ref{fig:Nk_xis} show the relationship between $\xi_s$ and $N_k$ by fixing $\lambda_S=10^{-13}$, and the variation of $\xi_S$ with respect to $N_k$ is remarkably little. However, the value of $\xi_S$ does not bring any impact on the constraints of preheating on inflation.

\begin{figure}[!h]
\begin{center}
\includegraphics[width=0.55\textwidth]{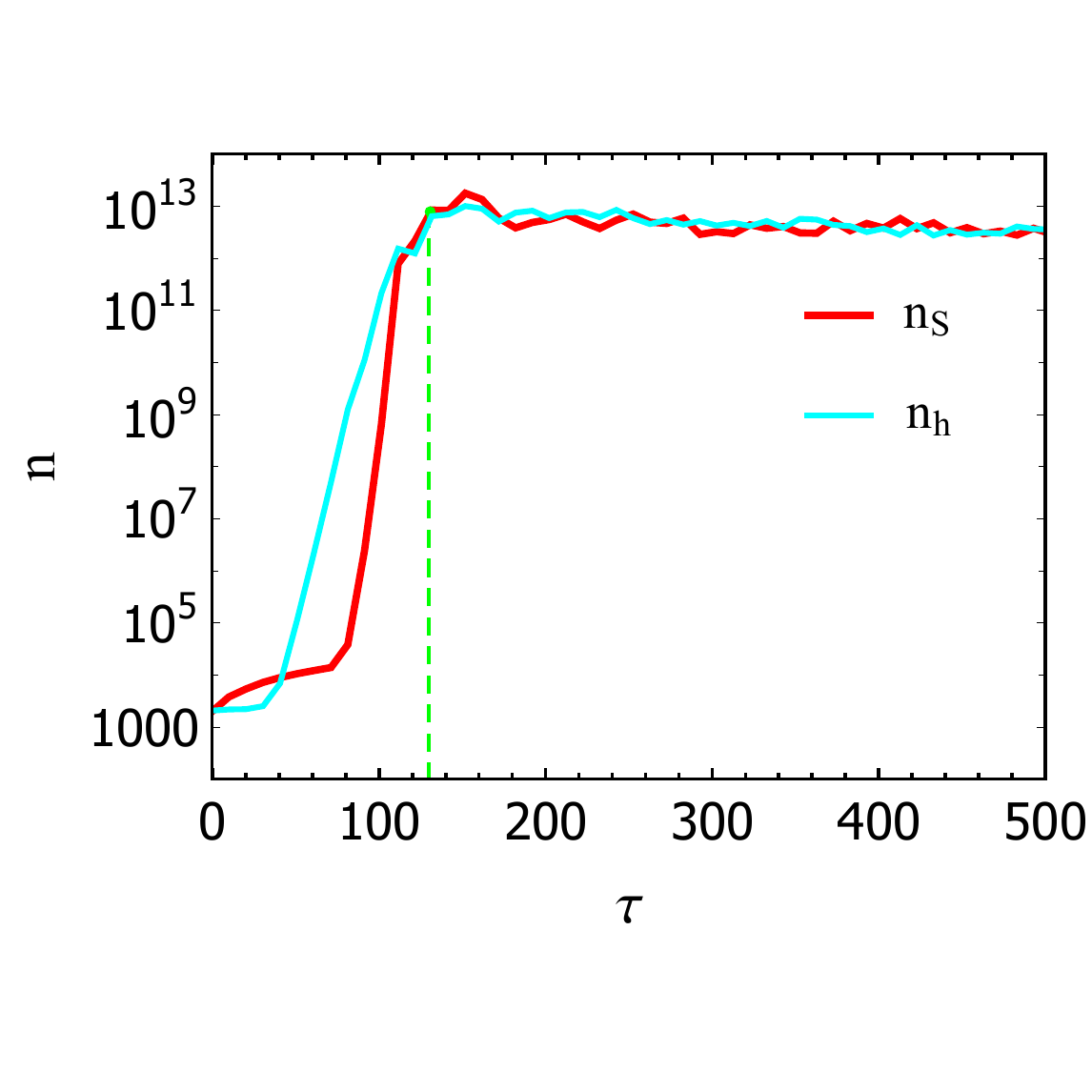}
\caption{The evolution of number density of S- and h- particles, where the red and blue lines represent S- and h- particle number densities respectively, and the green dashed line represents the end point of preheating.}
\label{fig:nt}
\end{center}
\end{figure}

The evolution of the number density of S- and h- particles are shown in Fig.~\ref{fig:nt} by applying the LATTICEEASY simulation. Note that the Higgs field indeed begins to emerge as the inflaton starts its decay process. However, it is crucial to point out that the decay of the inflaton does not strictly start immediately after inflation ends. This implies that particle production might be observable even before the end of inflation.~\cite{PIL}
At the beginning of evolution, particle number density grows exponentially. When the conformal time $\tau$ ($d\tau=\frac{\sqrt{\lambda_S}S_0 }{a}dt$, with $S_0=S(\tau=0)$) is about $130$, both particles almost reach the maximum number density, and then slowly change, which indicates that the evolution of the universe transitioned from preheating to reheating.

The physics behind this phenomenon can be explained as follows:
when the slow-rolling condition is upset, the inflaton will soon tumble to the lowest point of potential and oscillate periodically around the lowest point,
which causes the equation of motion of the h-field to become a Lame equation.

Parameter resonance for the h-field will emerge in some scenarios, which further bring an increase exponentially for the h-particle number, while the h-particle decay produces a significant backreaction for S-particles, and the S-particle number density also increases exponentially.

When the conformal time is about $130$, the amplitude of the S-field oscillation decreases, which leads to the weakening or even disappearance of the parametric resonance effect of the h-field, so that the h-particle number density no longer increases any more, and at the same conformal time, the backreaction effect also disappears, and the S-particle number density does not increase.

\begin{figure}[!htp]
\begin{center}
\includegraphics[width=0.47\textwidth]{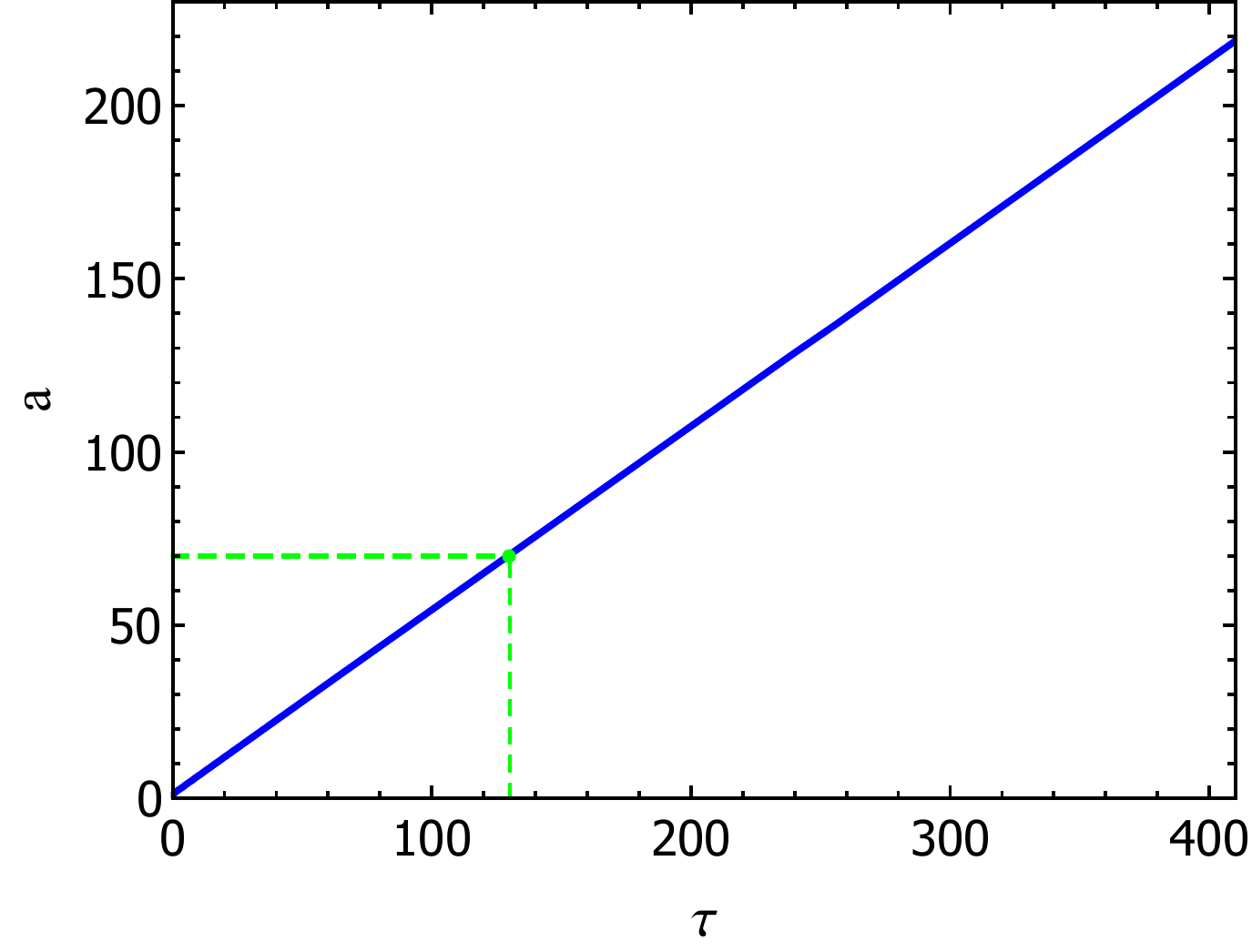}
\includegraphics[width=0.47\textwidth]{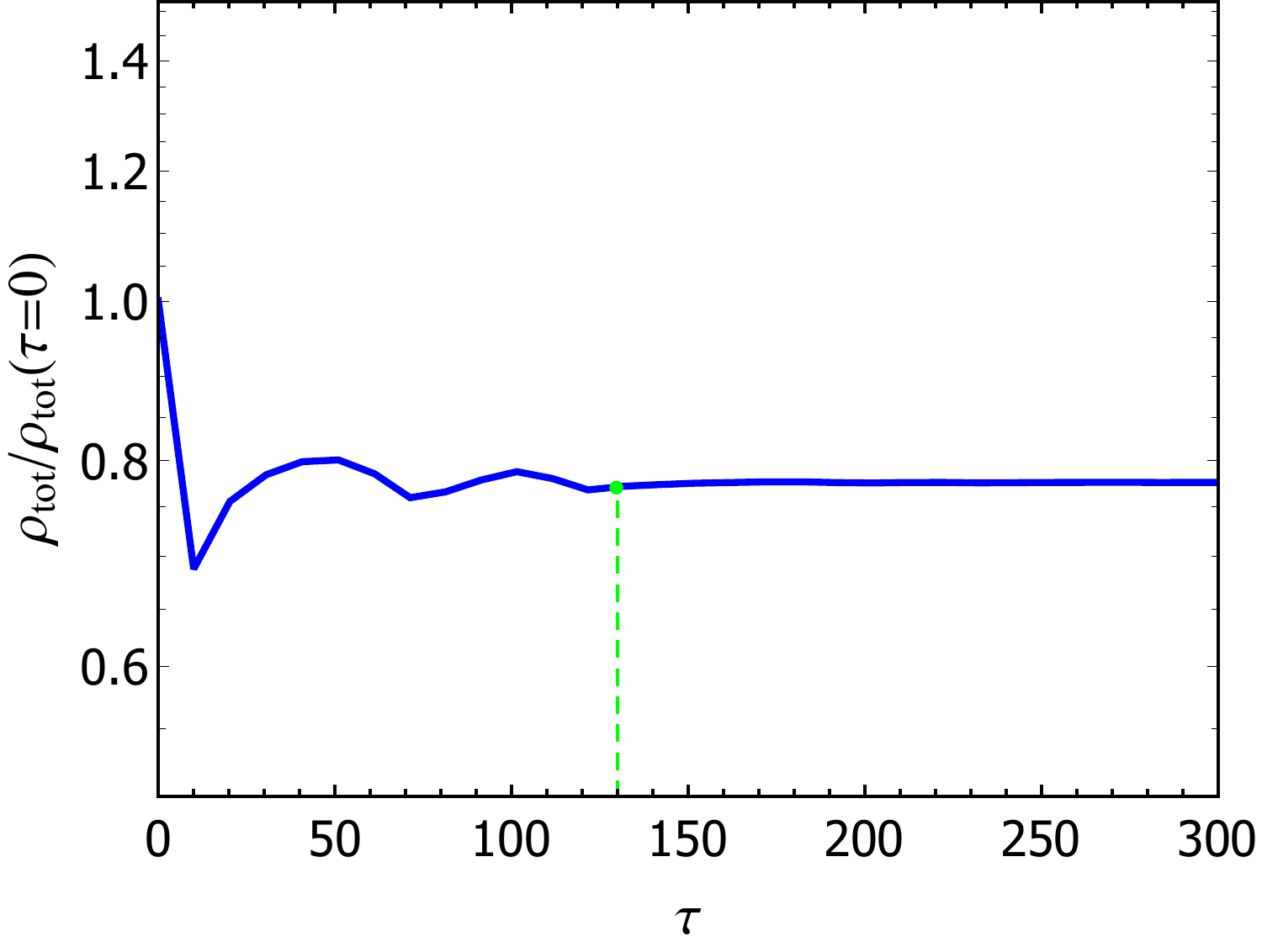}
\caption{The evolution of the scale factor and the total energy density for the left and right panels, respectively.}
\label{fig:atrhot}
\end{center}
\end{figure}
The variation of the scale factor over conformal time is shown on the left of Fig.~\ref{fig:atrhot}. The end conformal time of the preheating obtained by the simulation of particle number density evolution is about $130$, and the corresponding scale factor is $a_{end}\approx70$, while the value of the scale factor at the beginning of the preheating is $a_{star}=1$. Thus, one can calculate that the preheating e-folding number $N_{pre}=ln[a_{end}/a_{star}]\approx4.25$.

The evolution of normalized energy density $\rho_{tot}(\tau) / \rho_{tot}(\tau=0)$ with respect to conformal time $\tau$ is shown in the right of Fig.~\ref{fig:atrhot}, where the green dot is the ended energy of preheating according to the left of Fig.~\ref{fig:atrhot}.
Note that, $\rho_{tot}(\tau=0) = \rho_{end}$ and $\rho_{tot}(\tau=130) = \rho_{pre}$. Therefore, at $\tau=130$, the y-label of Fig.~\ref{fig:atrhot} is represented as $\rho_{tot}(\tau=130) / \rho_{tot}(\tau=0) \approx  0.77$, i.e., $\rho_{pre} / \rho_{end} = \gamma \approx 0.77$.
This simulation result is quite different from the assumption ($\gamma\approx10^3$ or $10^6$) in Ref.~\cite{ElBourakadi:2021nyb,ElBourakadi:2022lqf}.

\section{Numerical analysis and discussion}\label{sec:Num}

\begin{figure}[!htp]
\begin{center}
\includegraphics[width=0.55\textwidth]{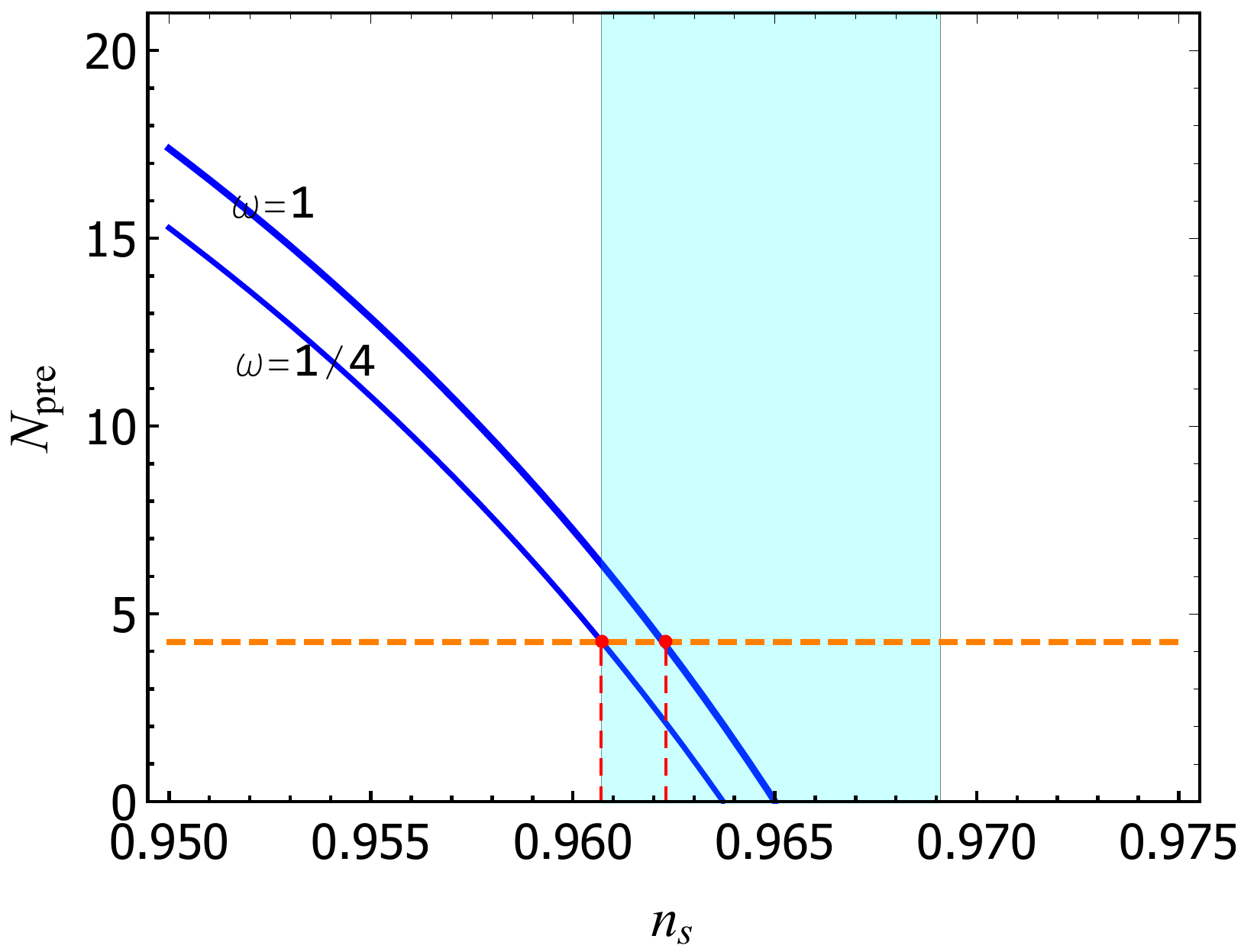}
\caption{The relation of scalar spectral index $n_s$ and e-folding number of inflation $N_k$, where the cyan shaded is the feasible area of the Planck limit~\cite{Planck:2018jri}, the orange dashed line is the $N_{pre}$ value obtained by LATTICEEASY simulation, the blue lines are calculated by the Eq.~(\ref{pre}), where the potential derived from the scalar inflation model and the energy ratio $\gamma$ are obtained by LATTICEEASY simulation.}
\label{fig:nsNpre}
\end{center}
\end{figure}

Fig.~\ref{fig:nsNpre} shows the relationship between the e-folding number of preheating $N_{pre}$ and scalar spectral index $n_s$. It indicates $N_{pre}$ increases with the state parameter $\omega$. After combining the prediction of $n_s$ from Planck with the LATTICEEASY simulation, the model has a feasible parameter space. The feasible range of $\omega$ is $1/4$ to $1$, and the corresponding $n_s$ range is $[0.9607, 0.9623]$.

\begin{figure}[!htp]
\begin{center}
\includegraphics[width=0.6 \textwidth]{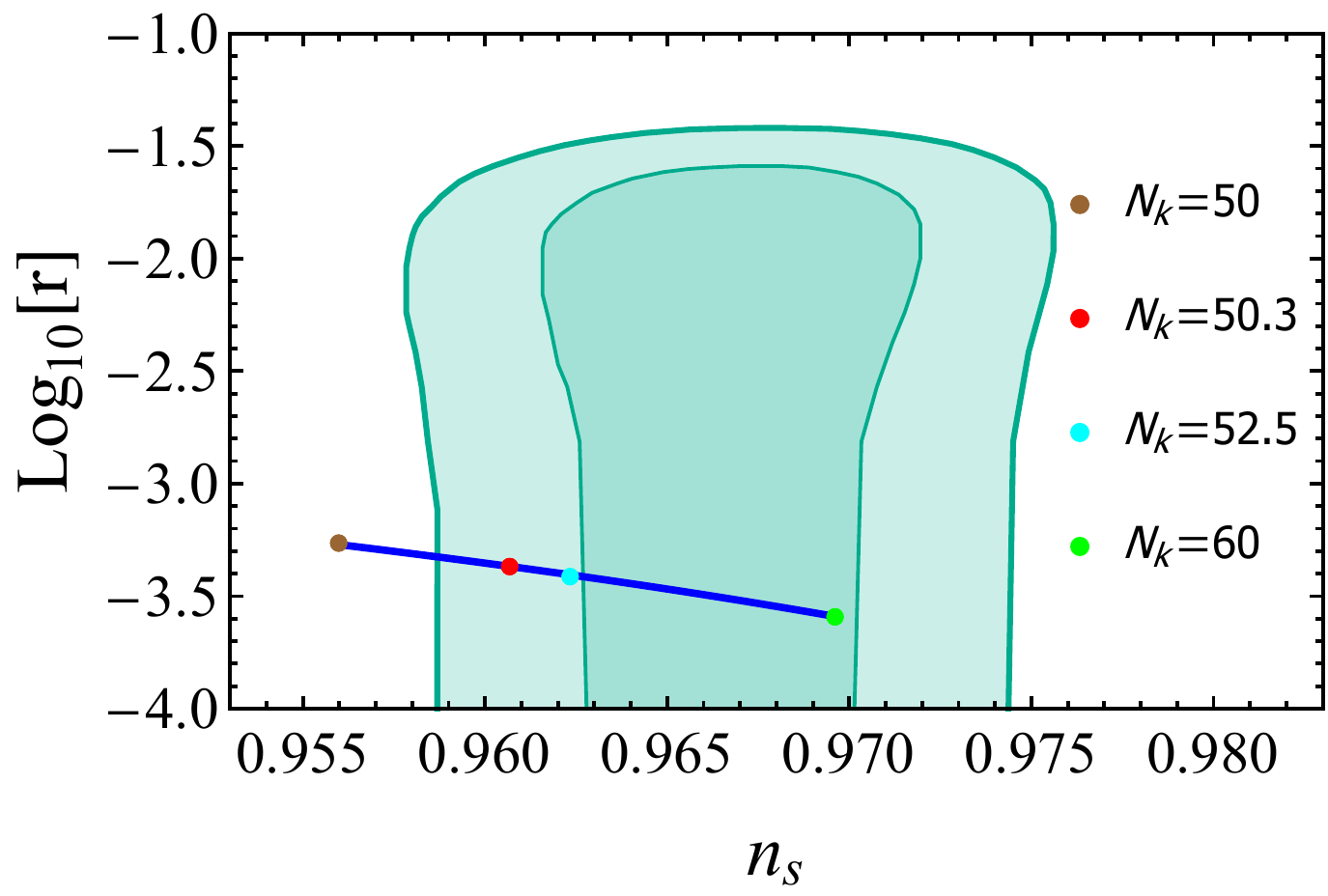}
\caption{The relation between scalar spectral index $n_s$ and Tensor-to-scalar power ratio $r$, where the blue line is our theoretical prediction for $N_k$ from $50$ to $60$, the green area are the Planck limits, and the brown, red, cyan and green points correspond to $N_k=50, 50.3, 52.5$ and $60$, respectively. }
\label{fig:nsr}
\end{center}
\end{figure}

The relation between $n_s$ and $r$ is shown in Fig.~\ref{fig:nsr}. Note that, the blue line between red point and cyan point is the feasible space of the LATTICEEASY simulation prediction obtained from Fig.~\ref{fig:nsNpre}. Therefore, the feasible parameter range of $r$ can be obtained, i.e., $r\thicksim[ 3.9\times10^{-4}, 4.3\times10^{-4}]$.

To illuminate the, we collect the fixed couplings of the scalar inflation model and the LATTICEEASY predictions in Table.~\ref{para}. With the same strategy, one can test other model parameters.

\begin{table}
\centering
\caption{LATTICEEASY simulation predicts with $\lambda_S=10^{-13}$, $\lambda_{Sh}=2\times10^{-11}$ and $\lambda_{h}=8\times10^{-12}$.}
\label{para}
\begin{tabular}{|c|c|c|c|c|c|}
  \hline
$\gamma$ & $N_{pre}$ & $w$ & $n_s$ & $N_k$ & $r\times 10^{-4}$ \\
  \hline
$0.77$ & $4.25$ & $[1/4,1]$ & $[0.9607,0.9623]$ & $[50.3,52.5]$ & $[ 3.9, 4.3]$ \\
  \hline
\end{tabular}
\end{table}

\section{Summary}\label{sec:Sum}
In this paper, we focus on studying a non-minimum coupled real scalar inflation model using the preheating phenomenon simulated by LATTICEEASY. During the inflation, the S-field plays a crucial role in driving the expansion of the universe, with its quartic term dominating the dynamics of inflation. Interestingly, we find that the evolutionary behavior of the inflationary potential remains unaffected by the coupling coefficient of the model. Furthermore, the predictions of important quantities derived from the model are also independent of this coupling coefficient.

Consequently, the tensor-to-scalar power ratio $r$ and scalar spectral index $n_s$ predicted by the Planck observations do not impose constraints on the coefficients of the model. However, by leveraging the relationship between preheating and inflation, the preheating process simulated using LATTICEEASY provides a valuable avenue to address this issue effectively.

We specifically investigate the constraint of inflation using the simulated preheating process, taking the couplings $\lambda_S=10^{-13}$, $\lambda_{Sh}=2\times10^{-11}$, and $\lambda_{h}=8\times10^{-12}$ as an illustrative example. By employing LATTICEEASY, we accurately reproduce the preheating dynamics within the scalar inflation model. With the introduction of LATTICEEASY, we have reproduced the preheating process of the scalar inflation model and obtained the particle number density evolution during the preheating process, and further deduced the end conformal time value of preheating. At the same conformal time, the e-folding number of preheating $N_{pre}$ and the energy ratio $\gamma$ are deduced by combining the simulated evolution of the scale factor and energy density, respectively.

By exploiting the relationship between preheating and inflation, we present the variation of $N_{pre}$ with $n_s$ in Fig.~\ref{fig:nsNpre}, enabling us to deduce the range of $w$ and $n_s$. Further, by exploring the connections between $n_s$ and $N_k$, as well as $N_k$ and $r$, we can obtain constraints on the ranges of $N_k$ and $r$. This implies that the preheating simulations conducted with LATTICEEASY can effectively predict $n_s$, $r$, and $N_k$, allowing for restrictions on scalar inflation models. Furthermore, this strategy can be extended to effectively constrain models involving inflaton dark matter model, as well as the interplay between inflation and dark matter model.

\hspace{2cm}

\acknowledgments

Wei Cheng was supported by Chongqing Natural Science Foundation project under Grant No. CSTB2022NSCQ-MSX0432, by Science and Technology Research Project of Chongqing Education Commission under Grant No. KJQN202200621, and by Chongqing Human Resources and Social Security Administration Program under Grants No. D63012022005.
Ruiyu Zhou was supported by Chongqing Natural Science Foundation project under Grant No. CSTB2022NSCQ-MSX0534.
This work was supported in part by the Fundamental Research Funds for the Central Universities under Grant No. 2021CDJQY-011, and by the National Natural Science Foundation of China under Grant No. 12147102.

\appendix

%\section{RGE}\label{app:rates}

\bibliographystyle{arxivref}

\begin{thebibliography}{999}

\bibitem{Guth:1980zm}
A.~H.~Guth, The Inflationary Universe: A Possible Solution to the Horizon and Flatness Problems, \href{https://doi.org/10.1103/PhysRevD.23.347}{Phys. Rev. D \textbf{23}, 347-356 (1981)}.

\bibitem{Linde:1981mu}
A.~D.~Linde, A New Inflationary Universe Scenario: A Possible Solution of the Horizon, Flatness, Homogeneity, Isotropy and Primordial Monopole Problems, \href{https://doi.org/10.1016/0370-2693(82)91219-9}{Phys. Lett. B \textbf{108}, 389-393 (1982)}.

%\cite{Planck:2015bpv}
\bibitem{Planck:2015bpv}
N.~Aghanim \textit{et al.} [Planck], Planck 2015 results. XI. CMB power spectra, likelihoods, and robustness of parameters, \href{https://doi.org/10.1051/0004-6361/201526926}{Astron. Astrophys. \textbf{594}, A11 (2016)}.

\bibitem{Planck:2018jri}
Y.~Akrami \textit{et al.} [Planck], {Planck 2018 results. X. Constraints on inflation},
\href{https://doi.org/10.1051/0004-6361/201833887}{Astron. Astrophys. \textbf{641}, A10 (2020)}.


\bibitem{Martin:2013tda}
J.~Martin, C.~Ringeval and V.~Vennin, {Encyclop\ae{}dia Inflationaris},
\href{https://doi.org/10.1016/j.dark.2014.01.003}{Phys. Dark Univ. \textbf{5-6}, 75-235 (2014)}.

\bibitem{Martin:2006rs}
J.~Martin and C.~Ringeval, {Inflation after WMAP3: Confronting the Slow-Roll and Exact Power Spectra to CMB Data},
\href{https://doi.org/10.1088/1475-7516/2006/08/009}{JCAP \textbf{08}, 009 (2006)}.

\bibitem{Martin:2010kz}
J.~Martin and C.~Ringeval,{First CMB Constraints on the Inflationary Reheating Temperature},
\href{https://doi.org/10.1103/PhysRevD.82.023511}{Phys. Rev. D \textbf{82}, 023511 (2010)}.

\bibitem{Dai:2014jja}
L.~Dai, M.~Kamionkowski and J.~Wang, {Reheating constraints to inflationary models},
\href{https://doi.org/10.1103/PhysRevLett.113.041302}{Phys. Rev. Lett. \textbf{113} (2014), 041302}.


\bibitem{Starobinsky:1980te}
A.~A.~Starobinsky, {A New Type of Isotropic Cosmological Models Without Singularity},
\href{https://doi.org/10.1016/0370-2693(80)90670-X}{Phys. Lett. B \textbf{91}, 99-102 (1980)}.

\bibitem{Jin:2020tmm}
G.~Jin, C.~Fu, P.~Wu and H.~Yu, {Production of gravitational waves during preheating in the Starobinsky inflationary model},
\href{https://doi.org/10.1140/epjc/s10052-020-8061-0}{Eur. Phys. J. C \textbf{80}, 491 (2020)}.

\bibitem{Mantziris:2022fuu}
A.~Mantziris, T.~Markkanen and A.~Rajantie, {The effective Higgs potential and vacuum decay in Starobinsky inflation},
\href{https://doi.org/10.1088/1475-7516/2022/10/073}{JCAP \textbf{10}, 073 (2022)}.


%
\bibitem{Bezrukov:2007ep}
F.~L.~Bezrukov and M.~Shaposhnikov, {The Standard Model Higgs boson as the inflaton},
\href{https://doi.org/10.1016/j.physletb.2007.11.072}{Phys. Lett. B \textbf{659}, 703-706 (2008)}.

\bibitem{Gundhi:2018wyz}
A.~Gundhi and C.~F.~Steinwachs, {Scalaron-Higgs inflation},
\href{https://doi.org/10.1016/j.nuclphysb.2020.114989}{Nucl. Phys. B \textbf{954}, 114989 (2020)}.

\bibitem{Antoniadis:2021axu}
I.~Antoniadis, A.~Guillen and K.~Tamvakis, {Ultraviolet behaviour of Higgs inflation models},
\href{https://doi.org/10.1007/JHEP08(2021)018}{JHEP \textbf{08}, 018 (2021)}.

\bibitem{Liu:2022myp}
Y.~Liu, {Higgs inflation and scalar weak gravity conjecture},
\href{https://doi.org/10.1140/epjc/s10052-022-10993-8}{Eur. Phys. J. C \textbf{82}, 1052 (2022)}.

\bibitem{Barrie:2021mwi}
N.~D.~Barrie, C.~Han and H.~Murayama, {Affleck-Dine Leptogenesis from Higgs Inflation},
\href{https://doi.org/10.1103/PhysRevLett.128.141801}{Phys. Rev. Lett. \textbf{128}, 141801 (2022)}.



\bibitem{Freese:1990rb}
K.~Freese, J.~A.~Frieman and A.~V.~Olinto, {Natural inflation with pseudo - Nambu-Goldstone bosons},
\href{https://doi.org/10.1103/PhysRevLett.65.3233}{Phys. Rev. Lett. \textbf{65}, 3233-3236 (1990)}.


\bibitem{Odintsov:2019mlf}
S.~D.~Odintsov and V.~K.~Oikonomou, {$f(R)$ Gravity Inflation with String-Corrected Axion Dark Matter},
\href{https://doi.org/10.1103/PhysRevD.104.063010}{Phys. Rev. D \textbf{99}, 064049 (2019)}.

\bibitem{Cheng:2021qmc}
W.~Cheng, L.~Bian and Y.~F.~Zhou, {Axionlike particle inflation and dark matter},
\href{https://doi.org/10.1103/PhysRevD.104.063010}{Phys. Rev. D \textbf{104}, 063010 (2021)}.

	
\bibitem{Zhang:2018wbn}
N.~Zhang, Y.~B.~Wu, J.~W.~Lu, C.~W.~Sun, L.~J.~Shou and H.~Z.~Xu, {Constraints on the generalized natural inflation after Planck 2018},
\href{https://doi.org/10.1088/1674-1137/44/9/095107}{Chin. Phys. C \textbf{44}, 095107 (2020)}.


%\cite{Linde:1983psb}
\bibitem{Linde:1983psb}
A.~D.~Linde, {PRIMORDIAL INFLATION WITHOUT PRIMORDIAL MONOPOLES},
\href{https://doi.org/10.1016/0370-2693(83)90316-7}{Phys. Lett. B \textbf{132}, 317-320 (1983)}.

\bibitem{Bostan:2019wsd}
N.~Bostan, {Quadratic, Higgs and hilltop potentials in the Palatini gravity},
\href{https://doi.org/10.1088/1572-9494/ab7ecb}{Commun. Theor. Phys. \textbf{72}, 085401 (2020)}.


\bibitem{Kallosh:2019jnl}
R.~Kallosh and A.~Linde, {On hilltop and brane inflation after Planck},
\href{https://doi.org/10.1088/1475-7516/2019/09/030}{JCAP \textbf{09}, 030 (2019)}.


\bibitem{Choubey:2017hsq}
S.~Choubey and A.~Kumar, {Inflation and Dark Matter in the Inert Doublet Model},
\href{https://doi.org/10.1007/JHEP11(2017)080}{JHEP \textbf{11}, 080 (2017)}.

\bibitem{Lebedev:2021zdh}
O.~Lebedev and J.~H.~Yoon, {Challenges for inflaton dark matter},
\href{https://doi.org/10.1016/j.physletb.2021.136614}{Phys. Lett. B \textbf{821}, 136614 (2021)}.

\bibitem{SBudhi:2019vln}
R.~H.~S.Budhi, {Inflation due to non-minimal coupling of f(R) gravity to a scalar field},
\href{https://doi.org/10.1088/1742-6596/1127/1/012018}{J. Phys. Conf. Ser. \textbf{1127}, 012018 (2019)}.

\bibitem{Cheng:2022hcm}
W.~Cheng, X.~Liu and R.~Zhou, {Non-minimal Coupling Inflation and Dark Matter under the $\mathbb{Z}_{3}$ Symmetry},
\href{https://arxiv.org/abs/2206.12624}{[arXiv:2206.12624 [hep-ph]].}

\bibitem{Albrecht:1982mp}
A.~Albrecht, P.~J.~Steinhardt, M.~S.~Turner and F.~Wilczek, {Reheating an Inflationary Universe},
\href{https://doi.org/10.1103/PhysRevLett.48.1437}{Phys. Rev. Lett. \textbf{48}, 1437 (1982)}.


\bibitem{Kofman:1994rk}
L.~Kofman, A.~D.~Linde and A.~A.~Starobinsky, {Reheating after inflation},
\href{https://doi.org/10.1103/PhysRevLett.73.3195}{Phys. Rev. Lett. \textbf{73}, 3195-3198 (1994)}.


\bibitem{Kofman:1997yn}
L.~Kofman, A.~D.~Linde and A.~A.~Starobinsky, {Towards the theory of reheating after inflation},
\href{https://doi.org/10.1103/PhysRevD.56.3258}{Phys. Rev. D \textbf{56}, 3258-3295 (1997)}.


\bibitem{Lozanov:2019jxc}
K.~D.~Lozanov, {Lectures on Reheating after Inflation},
\href{https://arxiv.org/pdf/1907.04402}{[arXiv:1907.04402 [astro-ph.CO]].}



\bibitem{Saha:2020bis}
P.~Saha, S.~Anand and L.~Sriramkumar, {Accounting for the time evolution of the equation of state parameter during reheating},
\href{https://doi.org/10.1103/PhysRevD.102.103511}{Phys. Rev. D \textbf{102}, 103511 (2020)}.


\bibitem{Cook:2015vqa}
J.~L.~Cook, E.~Dimastrogiovanni, D.~A.~Easson and L.~M.~Krauss, {Reheating predictions in single field inflation},
\href{https://doi.org/10.1088/1475-7516/2015/04/047}{JCAP \textbf{04}, 047 (2015)}.


\bibitem{Greene:1997fu}
P.~B.~Greene, L.~Kofman, A.~D.~Linde and A.~A.~Starobinsky, {Structure of resonance in preheating after inflation},
\href{https://doi.org/10.1103/PhysRevD.56.6175}{Phys. Rev. D \textbf{56}, 6175-6192 (1997)}.

\bibitem{Dolgov:1989us}
A.~D.~Dolgov and D.~P.~Kirilova,{ON PARTICLE CREATION BY A TIME DEPENDENT SCALAR FIELD},
\href{https://inspirehep.net/literature/281489}{Sov. J. Nucl. Phys. \textbf{51}, 172-177 (1990)}.

\bibitem{Traschen:1990sw}
J.~H.~Traschen and R.~H.~Brandenberger, {Particle Production During Out-of-equilibrium Phase Transitions},
\href{https://doi.org/10.1103/PhysRevD.42.2491}{Phys. Rev. D \textbf{42}, 2491-2504 (1990)}.

\bibitem{Shtanov:1994ce}
Y.~Shtanov, J.~H.~Traschen and R.~H.~Brandenberger, {Universe reheating after inflation},
\href{https://doi.org/10.1103/PhysRevD.51.5438}{Phys. Rev. D \textbf{51}, 5438-5455 (1995)}.

\bibitem{ElBourakadi:2021nyb}
K.~El Bourakadi, M.~Ferricha-Alami, H.~Filali, Z.~Sakhi and M.~Bennai, {Gravitational waves from preheating in Gauss\textendash{}Bonnet inflation}, \href{https://doi.org/10.1140/epjc/s10052-021-09946-4}{Eur. Phys. J. C \textbf{81}, 1144 (2021)}.

\bibitem{ElBourakadi:2022lqf}
K.~El Bourakadi, Z.~Sakhi and M.~Bennai, {Preheating constraints in \ensuremath{\alpha}-attractor inflation and gravitational waves production},
\href{https://doi.org/10.1142/S0217751X22501172}{Int. J. Mod. Phys. A \textbf{37}, 2250117 (2022)}.

%\cite{Cai:2021yvq}
\bibitem{Cai:2021yvq}
Y.~F.~Cai, J.~Jiang, M.~Sasaki, V.~Vardanyan and Z.~Zhou, {Beating the Lyth Bound by Parametric Resonance during Inflation},
\href{https://doi.org/10.1103/PhysRevLett.127.251301}{Phys. Rev. Lett. \textbf{127}, 251301 (2021)}


\bibitem{Felder:2000hj}
G.~N.~Felder, J.~Garcia-Bellido, P.~B.~Greene, L.~Kofman, A.~D.~Linde and I.~Tkachev, {Dynamics of symmetry breaking and tachyonic preheating},
\href{https://doi.org/10.1103/PhysRevLett.87.011601}{Phys. Rev. Lett. \textbf{87}, 011601 (2001)}.


\bibitem{Enqvist:2005qu}
K.~Enqvist, A.~Jokinen, A.~Mazumdar, T.~Multamaki and A.~Vaihkonen, {Non-Gaussianity from instant and tachyonic preheating},
\href{https://doi.org/10.1088/1475-7516/2005/03/010}{JCAP \textbf{03} (2005), 010}.

\bibitem{Dufaux:2008dn}
J.~F.~Dufaux, G.~Felder, L.~Kofman and O.~Navros, {Gravity Waves from Tachyonic Preheating after Hybrid Inflation},
\href{https://doi.org/10.1088/1475-7516/2009/03/001}{JCAP \textbf{03}, 001 (2009)}.


\bibitem{Felder:1998vq}
G.~N.~Felder, L.~Kofman and A.~D.~Linde, {Instant preheating},
\href{https://doi.org/10.1103/PhysRevD.59.123523}{Phys. Rev. D \textbf{59}, 123523 (1999)}.

\bibitem{Panda:2009ji}
S.~Panda, M.~Sami and I.~Thongkool, {Reheating the D-brane universe via instant preheating},
\href{https://doi.org/10.1103/PhysRevD.81.103506}{Phys. Rev. D \textbf{81}, 103506 (2010)}.


\bibitem{Dimopoulos:2017tud}
K.~Dimopoulos, L.~Donaldson Wood and C.~Owen, {Instant preheating in quintessential inflation with $\alpha$-attractors},
\href{https://doi.org/10.1103/PhysRevD.97.063525}{Phys. Rev. D \textbf{97}, 063525 (2018)}.

\bibitem{Cai:2016sdu}
Y.~F.~Cai, S.~Lin, J.~Liu and J.~R.~Sun, {Holographic Preheating},
\href{https://doi.org/10.3969/j.issn.0253-2778.2020.12.001}{J. Univ. Sci. Tech. China \textbf{50}, 1447-1452 (2020)}.

\bibitem{Cai:2016lqa}
Y.~F.~Cai, S.~Lin, J.~Liu and J.~R.~Sun, {Holographic Preheating: Quasi-Normal Modes and Holographic Renormalization},
\href{https://doi.org/10.3969/j.issn.0253-2778.2020.12.007}{J. Univ. Sci. Tech. China \textbf{50}, 1498-1506 (2020)}.


\bibitem{Felder:2000hq}
G.~N.~Felder and I.~Tkachev, {LATTICEEASY: A Program for lattice simulations of scalar fields in an expanding universe},
\href{https://doi.org/10.1016/j.cpc.2008.02.009}{Comput. Phys. Commun. \textbf{178}, 929-932 (2008)}.

%\bibitem{Bastero-Gil:2015lga}
%M.~Bastero-Gil, R.~Cerezo and J.~G.~Rosa, {Inflaton dark matter from incomplete decay},
%\href{https://doi.org/10.1103/PhysRevD.93.103531}{Phys. Rev. D \textbf{93}, 103531 (2016)}.
%
%
%\bibitem{Lebedev:2021xey}
%O.~Lebedev, {The Higgs portal to cosmology},
%\href{https://doi.org/10.1016/j.ppnp.2021.103881}{Prog. Part. Nucl. Phys. \textbf{120}, 103881 (2021)}.
%
%\bibitem{Cheng:2018axr}
%W.~Cheng and L.~Bian, {Higgs inflation and cosmological electroweak phase transition with N scalars in the post-Higgs era},
%\href{https://doi.org/10.1103/PhysRevD.99.035038}{Phys. Rev. D \textbf{99}, 035038 (2019)}.
%
%
%\bibitem{Cheng:2018ajh}
%W.~Cheng and L.~Bian, {From inflation to cosmological electroweak phase transition with a complex scalar singlet},
%\href{https://doi.org/10.1103/PhysRevD.98.023524}{Phys. Rev. D \textbf{98}, 023524 (2018)}.
%
%\bibitem{Belanger:2014bga}
%G.~B\'elanger, K.~Kannike, A.~Pukhov and M.~Raidal, {Minimal semi-annihilating $\mathbb{Z}_N$ scalar dark matter},
%\href{https://doi.org/10.1088/1475-7516/2014/06/021}{JCAP \textbf{06} 021, (2014)}.
%
\bibitem{Kaiser:2010ps}
D.~I.~Kaiser, {Conformal Transformations with Multiple Scalar Fields},
\href{https://doi.org/10.1103/PhysRevD.81.084044}{Phys. Rev. D \textbf{81}, 084044 (2010)}.

%
\bibitem{Zhou:2022ovp}
H.~Zhou, Q.~Yu, Y.~Pan, R.~Zhou and W.~Cheng, {Reheating constraints on modified single-field natural inflation models},
\href{https://doi.org/10.1140/epjc/s10052-022-10559-8}{Eur. Phys. J. C \textbf{82}, 588 (2022)}.
%
\bibitem{Planck:2015sxf}
P.~A.~R.~Ade \textit{et al.} [Planck],{Planck 2015 results. XX. Constraints on inflation},
\href{https://doi.org/10.1051/0004-6361/201525898}{Astron. Astrophys. \textbf{594}, A20 (2016).}

%
\bibitem{Koh:2018qcy}
S.~Koh, B.~H.~Lee and G.~Tumurtushaa, {Constraints on the reheating parameters after Gauss-Bonnet inflation from primordial gravitational waves},
\href{https://doi.org/10.1103/PhysRevD.98.103511}{Phys. Rev. D \textbf{98}, 103511 (2018)}.

\bibitem{Lebedev:2021ixj}
O.~Lebedev, T.~Nerdi, T.~Solomko and J.~H.~Yoon, Inflaton freeze-out, \href{https://doi.org/10.1103/PhysRevD.106.043537}{Phys. Rev. D \textbf{106}, 043537 (2022).}

\bibitem{Prokopec:1996rr}
T.~Prokopec and T.~G.~Roos, Lattice study of classical inflaton decay, \href{https://doi.org/10.1103/PhysRevD.55.3768}{Phys. Rev. D \textbf{55}, 3768-3775 (1997).}


\bibitem{Lebedev:2019ton}
O.~Lebedev and T.~Toma, Relativistic Freeze-in,
\href{https://inspirehep.net/undefined10.1016/j.physletb.2019.134961}{Phys. Lett. B \textbf{798}, 134961 (2019)}.


%\bibitem{Lebedev:2021zdh}
%O.~Lebedev and J.~H.~Yoon, ``Challenges for inflaton dark matter,''
%\href{https://inspirehep.net/undefined10.1016/j.physletb.2021.136614}{Phys. Lett. B \textbf{821}, 136614 (2021)}.

\bibitem{PIL}
F. Przybilla, Preheating in the lattice,
\href{https://javierrubioblog.files.wordpress.com/2016/09/report_fabian_przybilla.pdf}{(2016).}


\end{thebibliography}

\end{document}